\documentstyle[aas2pp4]{article}

\def\pd{\partial}

\lefthead{Ruffolo}
\righthead{Particles near an Oblique Shock}

\begin{document}

\title{Transport and Acceleration of Energetic Charged Particles \\ 
   near an Oblique Shock}

\author{D. Ruffolo}
\affil{Department of Physics, Chulalongkorn University, Bangkok
10330, Thailand}

\begin{abstract}
We have developed a numerical simulation code that treats the
transport and acceleration of charged particles crossing an idealized
oblique, non-relativistic shock within the framework of pitch angle 
transport using a finite-difference method.  We consider two applications:
1) to study the steady-state acceleration of energetic particles at an 
oblique shock, and 2) to explain observed precursors of Forbush decreases 
of galactic cosmic rays before the arrival of an interplanetary shock 
induced by solar activity.  For the former, we find that there is
a jump in the particle intensity at the shock, which is stronger for
more oblique shocks.  Detailed pitch angle distributions are also
presented.  The simple model of a Forbush decrease explains the key
features of observed precursors, an enhanced diurnal anisotropy 
extending several mean free paths upstream of the shock and a
depletion of particles in a narrow loss cone at $\sim0.1$ mean
free path from the shock.  Such precursors have practical
applications for space weather prediction.

\end{abstract}


\section{Introduction}

An amazing variety of astrophysical phenomena can be attributed to
magnetohydrodynamic shocks, and in particular to 
their ability to accelerate charged particles to high energies
(\cite{be87}).
Shock acceleration is believed to account for the bulk of the galactic 
cosmic rays (\cite{a81}) and solar cosmic rays (\cite{lr86})
observed near the Earth.
There is a variety of mechanisms by which shocks can accelerate
particles, as indicated by direct observations within the solar
system.  Diffusive shock acceleration (\cite{k77,als77,b78}), 
based on the first-order Fermi acceleration mechanism
(\cite{f54,p58}), can accurately account for acceleration at the Earth's 
bow shock (\cite{e81,emp90}) and traveling interplanetary shocks
(\cite{l83,kea86}), whereas stochastic
acceleration, or second-order Fermi acceleration (\cite{f49}), 
is apparently the dominant mechanism in the vicinity of cometary bow 
shocks (\cite{ts89}). 

Several lines of evidence indicate the distributed
acceleration of energetic ions over a wide range of
heliolongitudes during gradual solar flare/coronal mass ejection
(CME) events (\cite{mgh84,lr86,r90,r97}), presumably due to
CME-driven shocks, though it is not clear whether the first- or 
second-order Fermi
mechanism is responsible for initial particle acceleration near the
Sun.  As CME shocks move outward from the Sun, they become
traveling interplanetary shocks, and as such a shock passes a detector,
intensity enhancements or anisotropy variations of energetic particles are
often recorded.  Evidence that such intensity enhancements can
represent true acceleration, not mere pile-up, comes from major
spectral changes (\cite{m72,rea97}).  At energies $\lesssim 1$ MeV, there is 
clear evidence that particles can be accelerated out of the solar wind
population (\cite{gea81}), presumably by diffusive shock acceleration.  
However, for energies $>10$ MeV, all observations to date of ionic
charge states near the times of interplanetary shock arrivals
are inconsistent with a solar wind origin, 
indicating that the ions were originally accelerated out of coronal 
material (\cite{bta96}).  In addition, Tan et al.\ (1989) 
provided compositional evidence 
that ions of $\sim1$ MeV observed at the time of shock passage represent 
the same population as ambient solar energetic particles.    
The above evidence indicates that particles are injected into the 
interplanetary medium while the shock is still near the Sun, and 
later this particle population may be further affected by the shock 
as it propagates through the inner heliosphere.  
This leads one to the question of how an oblique shock (i.e., a shock
in which the magnetic field is neither parallel nor perpendicular to
the shock normal) further accelerates
existing populations of energetic charged particles.

Note that close to an oblique shock,
the diffusion approximation does not provide an accurate description
of the spatial or directional distribution of energetic particles.  
The goal of the present work is to examine the spatial and pitch angle 
distribution that arises due to the transport and acceleration of
an existing particle population near an oblique shock.  
To the author's knowledge, this represents the first solution of the
pitch angle transport equation on both sides of an oblique,
non-relativistic shock for non-ultrarelativistic particles.  
(For ultrarelativistic particles at oblique shocks, for
which one can set $E\approx pc$, see \cite{kh89}; for parallel
shocks and non-relativistic particles, see \cite{ks89}.)
Therefore, we have started with
the simplest case of an oblique shock with a constant magnetic field
on either side in a medium with a spatially uniform scattering mean
free path (Figure 1).  
It is hoped that this will lay the groundwork for further studies
of particle acceleration integrated into the framework of a pitch angle 
transport equation that will consider other effects for particular types 
of shocks; these could include a realistic magnetic field configuration, a
spatially dependent scattering amplitude, a
self-consistent treatment of wave generation and pitch angle scattering, 
or other effects.

\placefigure{f:config}

The basic process of charged particle acceleration by planar, parallel 
shocks has been worked out in the diffusion approximation 
(e.g., \cite{k77,bo78}).  Decker \& Vlahos (1986) and Jokipii (1987) found 
that for oblique shocks the rate of particle acceleration increases 
with the shock angle, i.e., the angle between the magnetic field and the 
shock normal.  In addition, there is a 
characteristic length over which the particle distribution increases 
upstream of the shock, given by $D/u=v\lambda_\parallel/(3u)$, where $D$ 
is the coefficient of spatial diffusion (due to pitch angle scattering), 
$u$ is the fluid speed relative to the shock, $v$ is the particle speed, 
and $\lambda_\parallel$ is the spatial mean free path along the magnetic 
field.  In the inner heliosphere, $\lambda_\parallel$ is typically 0.08 
to 0.3 AU (\cite{p82}), and for $u\sim 0.002c$ and $v=0.1c$ 
($E=94$ MeV), this length scale 
is on the order of 17$\lambda_\parallel$, or 1.4 to 5 AU.  Thus a model with a 
constant diffusion coefficient does not 
explain the more localized increases observed as an interplanetary shock 
passes the Earth.  One explanation is that the diffusion coefficient 
may decrease near a shock due to increased magnetic turbulence generated 
by accelerated particles (e.g., \cite{mw68,l83,rbn96}).  On the other 
hand, the large changes in particle anisotropies that are sometimes
observed during shock passage (e.g., \cite{emy82,hea95})
indicate a breakdown in the diffusion approximation, so a 
detailed treatment should also consider the pitch angle distribution.  
Recently, the EPAM instrument on the ACE spacecraft has 
been able to measure complex pitch angle distributions close to
the time of shock passage (\cite{ah98}); models of shock acceleration should
attempt to reproduce such distributions.  Furthermore, changes in the
pitch angle distribution of relativistic ions (mainly galactic cosmic ray
protons) as an interplanetary shock approaches, representing
precursors to a Forbush decrease accompanying the shock
(\cite{f38,bh42,fl42}) can be measured by ground-based neutron monitors and 
could provide advance warning of approaching shocks (\cite{be97}), which can 
induce geomagnetic storms and affect satellites, communications, and 
power grids at Earth.  Clearly the implementation of such a warning 
system would rely on an accurate understanding of the pitch angle
distribution of energetic ions upstream of an interplanetary shock.

Kirk \& Schneider (1987a) presented 
an approximate analytic solution of a pitch angle transport equation 
for highly relativistic particles near a parallel shock discontinuity
(i.e., one in which the magnetic field is parallel to the shock
normal), giving insight into the variety of length scales corresponding to 
higher orders of anisotropy near the shock (the length scales also apply to
non-relativistic particles).
This work was extended to oblique shocks by Kirk \& Heavens (1989).
Monte Carlo techniques have provided important information on 
the pitch angle distribution near the shock (\cite{ks87b,nt95}),
the final spectrum (\cite{ks87b,bh91,lea94,nt95}),
the rate of particle acceleration vs.\ the magnetic field-shock angle
(\cite{dv86,lea94,nt95}), the injection 
of particles from a thermal population and its decreased efficiency
for larger magnetic field-shock angles (\cite{bej93}), compositional 
selection effects (\cite{eje81}), and the relative importance of
first- and second-order Fermi acceleration at shocks (\cite{ka94}).

Here we adopt a different approach by incorporating the treatment 
of an oblique shock discontinuity into a finite-difference numerical 
solution of a pitch angle transport equation.
Previously, numerical techniques have been developed to treat the
pitch angle dependent transport of energetic particles in the inner
heliosphere (\cite{nw79,s85,e87,r91,pb94}), 
recently including the effects of adiabatic deceleration and
convection (\cite{r95,h96,l97,kea98}), and these have been used to
analyze observations of solar energetic particles
(\cite{bea80,bep86,dwk90,kwh92,rky98})
and solar neutron decay particles (\cite{r91,drk96}).
Including the injection of particles from other sources,
such as a traveling interplanetary shock, has proven more difficult.
Previous authors have set the simulation boundary at
the shock and assumed an {\it ad hoc} particle injection function,
say, $Q$, finding the function that best fits the observed data for a given
event and energy range (\cite{hea95,l97,kw97}; Kallenrode 1997a, b).
In principle this allows one to examine the dependence of 
$Q$ on physical parameters of the shock, though there are a number
of such interrelated parameters, which complicates the interpretation
of the association between $Q$ and any individual parameter.
Our approach is in a sense the reverse: we examine what parameter
dependence should be expected based on certain physical processes.
Since diffusive shock acceleration is a manifestation of particle 
transport in the vicinity of the shock, we include both sides of the
shock discontinuity in the calculation, with special treatment of
particles crossing the shock based on the conservation of
the magnetic moment, to examine both the transport and further acceleration 
of energetic charged particles near the shock.  We are able to examine the 
effects of the shock-field angle and the form of the pitch angle scattering
on the steady-state particle distribution in space and pitch angle
for an idealized model of an oblique shock.  As a further application
of our method, we apply it to explain the observed precursors of
Forbush decreases before the arrival of an interplanetary shock.

\section{Methodology}

\subsection{Overview}
Before describing the methodology in detail, we present a brief 
overview of shock acceleration.  The process of ``diffusive shock 
acceleration'' refers to the acceleration of charged particles as they 
are repeatedly scattered back and forth across a shock front, which
for the present purposes is a discontinuity in the fluid speed and
magnetic field strength.  This process does not require the diffusion 
approximation and can be examined by lower-level descriptions
in terms of pitch angle scattering (\cite{ks87a}) 
or particle orbits in a disordered magnetic field (\cite{dv85}).

Originally the literature
discussed two mechanisms of acceleration at shocks:

1. In the first-order
Fermi acceleration mechanism (\cite{f54,p58}), particles gain energy
when scattering off of converging magnetic field irregularities.
At a shock, this occurs because the field irregularities are
convected with the fluid flow (in the r\'egime in which the Alfv\'en
speed is smaller than the fluid speed relative to the shock), 
and the upstream fluid flows toward
the shock faster than the downstream fluid flows away.  
In this r\'egime, scattering tends to isotropize the momentum 
while preserving its magnitude in the local fluid frame.  
When a particle heads toward the shock (in the local fluid frame)
in either direction, the frame transformation 
leads to a higher momentum in the new local fluid frame, which is
conserved until the particle encounters the shock again.  After two such
crossings of the shock, the particle has a higher momentum in
its original reference frame.

2. In the shock drift mechanism (\cite{s63}),
particles drift along an oblique shock front due to the 
sharp gradient in the magnetic field, and this drift is along the
direction of the electric field so that particles can gain a
substantial amount of energy in one encounter with the shock.

More recently, it has been shown that the distinction between these
two mechanisms vanishes in the de Hoffmann-Teller (shock) frame (\cite{dt50}) 
where the electric field is zero.  The entire energy change due to
mechanisms 1 and 2 is accounted for by 
transforming the particle momentum from the local fluid frame into the
shock frame, considering the energy-conserving shock encounter in the 
shock frame, and then transforming the momentum into the new local 
fluid frame (\cite{dec83}).
[Note, however, that there is still a lateral drift along the shock, which
is important, e.g., for the acceleration of 
anomalous cosmic rays at the solar wind termination shock
(\cite{pej81,csw86}).] For an oblique shock, 
the term diffusive shock acceleration is now generally used to refer to 
this unified description of particle acceleration that includes both
mechanisms 1 and 2.

The basic mechanism of particle acceleration at an oblique shock, as
described above, includes two processes: a) frame transformations, as in
first-order Fermi acceleration, and b) a change in pitch angle as the
particle encounters the shock (which does not occur for a parallel
shock).  Processes a) and b) are related to parallel and perpendicular 
changes in the fluid velocity, respectively.  Thus they are
directly analogous to the two components of adiabatic deceleration in
the solar wind (e.g., \cite{wg79,r95}), which are associated with the
divergence of the wind parallel and perpendicular to the magnetic
field and have been termed ``an inverse Fermi effect'' and ``betatron
deceleration,'' respectively.  Just as betatron deceleration is an
effect of adiabatic focusing (also known as magnetic mirroring) when viewed 
in the local fluid frame (\cite{r95}), which in turn represents the 
conservation of the magnetic moment 
$p_\perp^2/(2meB)$, for the case of shock acceleration the change in 
pitch angle as particles encounter the shock approximately conserves the 
magnetic moment as well (see \S 2.5).  If we assume that the magnetic
moment is actually conserved, then we do not need to concern
ourselves with the details of the particle orbit near the shock; we
treat the entire particle-shock interaction as a single event, and
consider whether a particle ultimately crosses or is reflected by the
shock, which is assumed to depend only on the initial pitch angle.

Figure 2 provides a schematic illustration (for a non-relativistic particle) 
of acceleration at an oblique shock.
[For an analogous illustration for adiabatic deceleration, see Figure 1 of 
Ruffolo (1995).]  The components of the particle velocity, $v_\parallel$ and 
$v_\perp$, refer to motion parallel and perpendicular to the magnetic field,
respectively (the pitch angle is the angle with respect to the
positive $v_\parallel$ axis).  Let us take the upstream direction to 
be to the right; then the upstream and downstream flow speeds with 
respect to the shock frame are negative (``U'' and ``D,'' 
respectively, in Figure 2).  A particle with $v_\parallel<0$ is
moving from the upstream side toward the downstream side.  The
velocity relative to point ``D'' is greater than that relative to
point ``U,'' so if the particle crosses to the downstream side, the 
magnitude of its velocity is greater in the new local fluid frame, and this
velocity magnitude is conserved by further scattering in that frame.
Similarly, a particle with $v_\parallel>0$ can cross from downstream to
upstream, and again the magnitude of the velocity increases in the
new local fluid frame.  A third type of shock encounter (for oblique
shocks) is when a particle from upstream ($v_\parallel<0$) reflects back 
upstream with $v_\parallel \rightarrow -v_\parallel$ in the shock frame; 
again, we see that the magnitude of the velocity increases in the fluid 
frame.

\placefigure{f:circle}

In addition to the frame transformations described above, i.e.,
standard first-order Fermi acceleration for parallel shocks, at
oblique shocks there is the additional process of a change in pitch
angle as a particle crosses the shock.
At first glance it is not immediately obvious why this process should 
lead to acceleration as opposed to deceleration.  
To see this, consider 
again a particle with $v_\parallel<0$ crossing the shock from 
upstream to downstream (Figure 2).
In the shock frame the large-scale magnetic field configuration is static,
so the particle speed in this frame, $\sqrt{v_\parallel^2+v_\perp^2}$, 
is conserved.  If we assume approximate conservation of the magnetic
moment, then $v_\perp$ will increase, since the magnetic field is
stronger on the downstream side.  As can be seen in Figure 2, 
the speed relative to the downstream fluid (``D'') increases; the
dotted line indicates a curve of constant speed in the downstream
fluid frame.  Thus this process leads to an additional increase in 
energy in the new local fluid frame, beyond that due to the
frame transformation.  Similarly, a particle with $v_\parallel>0$ is moving
from downstream to upstream, and approximate conservation of the magnetic
moment implies a decrease in $v_\perp$.  Then the speed relative to the
upstream fluid (``U'') increases; the dashed line indicates a curve
of constant speed in the upstream fluid frame.  The change in pitch
angle can be quite large, so this can lead to a substantial
increase in the energy gain at oblique shocks.  There can also be a
significant effect on the pitch angle distribution, which makes
particle transport and acceleration at 
oblique shocks more difficult to model than that at parallel shocks.

\subsection{Transport Equation}
To our knowledge, this work 
represents the first treatment of oblique shock acceleration within a
finite difference simulation of pitch angle transport, so as a first 
step this work considers only a simple, planar, oblique shock with 
straight magnetic field lines on either side (Figure 1).  We consider 
only subluminal shocks so that we can work in the de Hoffmann-Teller 
reference frame (which we also refer to as the ``shock frame'')
in which the shock is stationary, the fluid flow is parallel to the magnetic 
field, and the electric field is zero (\cite{dt50}).

Ruffolo (1995) provided an equation of focused pitch angle transport 
(eq.\ [11] of that paper) that included solar wind effects, such
as adiabatic deceleration and convection, to first order in $u/c$ 
in an Archimedean spiral 
magnetic field.  When we consider a region of constant 
magnetic field on either side of the shock, that equation reduces to
\begin{eqnarray}
\frac{\pd F(t,\mu,z,p)}{\pd t} & = & -\frac{\pd}{\pd z}\mu vF(t,\mu,z,p)
 \nonumber \\
& & \mbox{} - \frac{\pd}{\pd z}\left(1-\mu^2\frac{v^2}{c^2}\right)
 uF(t,\mu,z,p) \nonumber \\
& & \mbox{} + \frac{\pd}{\pd\mu}\frac{\varphi(\mu)}{2}\frac{\pd}{\pd\mu}
 \left(1-\mu\frac{uv}{c^2}\right) \nonumber \\
& & \mbox{\ \ } \cdot F(t,\mu,z,p),
\label{eq:tr}
\end{eqnarray}

\begin{flushleft}    
\begin{tabular}{rp{6.5cm}}
where $F$ & $\equiv d^3N/(dzd\mu dp)$ is the density of \\
 & particles in a given magnetic flux tube, \\
$t$ & is the time in the shock frame, \\
$\mu$ & is the pitch angle cosine in the wind \\
 & frame, \\
$z$ & is the distance from the shock along the \\
 & magnetic field in the shock frame, \\
$p$ & is the particle momentum in the wind \\
 & frame, \\
$v$ & is the particle velocity in the wind frame, \\
$u$ & is the solar wind speed along the \\ 
 & magnetic field in the shock frame, and \\
$\varphi$ & is the pitch angle scattering coefficient.
\end{tabular}
\end{flushleft}

\noindent In equation~(\ref{eq:tr}), the first term on the right hand
side represents the effect of streaming, the second is for convection 
(including the relativistic correction for transforming the streaming
speed into the shock frame), and the third is for pitch angle scattering.
The particle density in a flux tube, $F$, is related to the phase 
space density, $f$, by $F=2\pi ap^2f$, where $a(z)\propto 1/B(z)$ is the 
cross-sectional area of the flux tube.  We use $F$ in the simulations,
following Ng and Wong (1979), because we can easily design the numerical
finite difference method to strictly conserve this quantity during streaming 
and convection, even when $a(z)$ and $B(z)$ are spatially varying.
The pitch angle scattering coefficient, $\varphi(\mu)$, is expressed as
\begin{equation}
\varphi(\mu)=A|\mu|^{q-1}(1-\mu^2)
\label{eq:phi}
\end{equation}
where $A$ is the scattering amplitude and $q$ controls the form of 
the scattering coefficient.  This expression, originally derived in
the context of quasi-linear theory (\cite{j71}), is adopted as a 
convenient and widely used parameterization.  In this work we will
consider $q=1$, for isotropic scattering, and $q=1.5$, which is in the
range of 1.3 to 1.7 inferred by Bieber et al.\ (1986) for interplanetary 
scattering.  Roughly speaking, $q>1$ implies a deficit in scattering
near a pitch angle of 90$^\circ$ ($\mu=0$); the effect of
such a deficit on shock acceleration has also been examined by Kirk
(1988).

\subsection{Eigenfunction Expansion}

In addition to our numerical solution of equation~(\ref{eq:tr})
(see \S2.6),
it is possible to find analytic solutions in the steady 
state if we restrict the $z$ domain to one side of the shock.  Setting 
$\pd F/\pd t=0$, we have
\begin{eqnarray}
\frac{\pd F}{\pd z} & = & \frac{A}{2[\mu v(1-\mu uv/c^2)+u]} \nonumber \\
 & & \cdot\frac{\pd}{\pd\mu} (1-\mu^2)|\mu|^{q-1} \frac{\pd}{\pd\mu} 
   \left(1-\mu\frac{uv}{c^2}\right)F.
\label{eq:ss1}
\end{eqnarray}
Note that $F$ is defined in a mixed frame, i.e., for $\mu$ and $p$
defined in the solar wind frame and $z$ and $t$ in the shock frame.
To first order in $u/c$, the analogous quantity defined in terms of
variables in the solar wind 
frame is given by $F_w=(1-\mu uv/c^2)F$ (\cite{wg79}), so we can simplify 
equation (\ref{eq:ss1}) to read
\begin{equation} 
\frac{\pd F_w}{\pd z}=\frac{A}{2[\mu v+ u]} \frac{\pd}{\pd\mu} 
  (1-\mu^2)|\mu|^{q-1} \frac{\pd F_w}{\pd\mu},
\end{equation}
again neglecting terms of order $(u/c)^2$.
For a given $p$, this can be solved by separation of variables, 
$F_w(\mu,z)=M(\mu)Z(z)$, yielding $Z\propto e^{kz}$ and 
\begin{equation}
\frac{\pd}{\pd\mu}(1-\mu^2)|\mu|^{q-1}\frac{\pd M}{\pd\mu}
  -\alpha\left(\mu+\frac{u}{v}\right)M = 0,
\label{eq:e}
\end{equation}
where $\alpha\equiv 2kv/A$ is an eigenvalue of the equation.  To avoid
divergence as $z\rightarrow\pm\infty$, we must have $k\leq0\ (\geq0)$
for $z>0\ (<0)$.
Following Kirk \& Schneider (1987a), who treated the case of $q=1$,
$M(\mu)$ can be expanded in terms of normalized Legendre 
polynomials.  For $q$ and $u/v$ values of interest, we truncated the 
expansion to $n$ terms and evaluated eigenvalues and eigenfunctions using the 
Mathematica software package (Wolfram Research, Inc.).
For $q=1$, $n=12$ was sufficient to obtain eigenvalues with a
relative accuracy of 10$^{-5}$.  For $q=1.5$, 
obtaining an accuracy of about 2\% for the first positive eigenvalue
required $n\approx 80$; for other eigenvalues fewer terms were
required.

\placefigure{f:eigen}

The eigenvalues $\alpha$ closest to zero (corresponding to long spatial
scales) are shown in Figure 3 for $q=1$ and 1.5 and a range of $u/v$
values.  For $u/v=0$, the numerical values of $\alpha$ are
0, $\pm$14.53, $\pm$42.05, $\pm$83.30, $\pm$138.3, \dots 
for $q=1$ and 0, $\pm$11.3, $\pm$33.0, $\pm$65.6, $\pm$109, \dots for $q=1.5$.
For a given $q$ the relative magnitudes of these eigenvalues are similar, but
not identical, to those of the Legendre polynomials, $\ell(\ell+1)$ for
$\ell=0,1,2,\dots$.  The eigenvalues shown in Figure 3 for
$q=1.5$ are about 20\% lower in magnitude than those for
$q=1$, with the exception that $\alpha_1$ is lower by a factor of about 2.4.
Aside from $\alpha_1$, the eigenvalues shown here are approximately given by
$\alpha_i=\alpha_i^0(1\pm\beta u/v)$, where $\alpha_i^0$ is the value
for $u/v=0$, ``$\pm$'' follows the sign of $\alpha_i^0$, 
and $\beta$ is a constant between 2.2 and 2.4.
Sample normalized eigenfunctions for these eigenvalues 
are shown in Figure \ref{f:eigenfunc} for
$q=1$ and $u/v=-0.075$, with particles traveling away from the shock
for $\mu<0$ downstream and $\mu>0$ upstream; each eigenfunction
$M(\mu)$ is plotted with the sign for which $\langle M\rangle>0$.  
In cases considered in this paper, $uv/c^2\ll 1$, so $F_w\approx F$
and $M$ essentially gives the angular dependence of $F$.

\placefigure{f:eigenfunc}

For $u/v>0$, there is a zero eigenvalue, $\alpha_0$, with a 
corresponding constant eigenfunction, and 
a first positive eigenvalue, $\alpha_1$, which tends to zero as
$u/v\rightarrow 0$.  For $q=1$, $\alpha_1\approx6u/v$ (with slight 
deviations for higher $u/v$), implying that $k\approx 3uA/v^2$.  Note that in
the diffusion approximation, the spatial diffusion coefficient is given by
the classic formula (Jokipii 1966, 1968, \cite{hw68}),
\begin{equation}
D = \frac{v^2}{2}\int_0^1\frac{(1-\mu^2)^2}{\varphi(\mu)}d\mu,
\label{eq:d}
\end{equation}
and for the form of $\varphi(\mu)$ specified in equation~(\ref{eq:phi}), we
have the well-known expression
\begin{equation}
D = \frac{v^2}{A}\frac{1}{(2-q)(4-q)}.
\end{equation}
For $q=1$ we have $D=v^2/(3A)$, and the
approximate $z$-dependence corresponding to the first positive eigenvalue
is $Z\propto e^{uz/D}$; such a dependence is also found for the case of
$q=1.5$.  Note that this solution only applies to the region where
$uz<0$, i.e., the upstream region.  In general, solutions
corresponding to positive (negative) $\alpha$ are valid in
the upstream (downstream) region, and the constant eigenfunction
corresponding to $\alpha_0=0$ is valid in either region.

The complete solution for $F$ in the steady state on either side of
the shock is a linear combination of the separable solutions.
Therefore, the solution far upstream of the shock is a superposition
of a constant and the above solution proportional to $e^{uz/D}$.
Since $|\alpha_{-1}|\gg \alpha_1$, the only valid solution far
downstream of the shock is a constant.
Such spatial behavior of $F$ at long distances from
the shock is in accordance with standard results
based on the diffusion approximation (e.g., \cite{k77}; 
see also \S 2.5), though there are slight deviations at higher $u/v$ 
(for $q=1$, $\alpha_1>6u/v$ by $\approx$2\% at $u/v=0.1$).
At shorter distances there may be contributions from solutions
corresponding to other eigenvalues.  In terms of the particle mean free path, 
\begin{equation}
\lambda = \frac{3D}{v} = \frac{v}{A}\frac{3}{(2-q)(4-q)},
\end{equation}
such solutions decay over a distance scale of 
\begin{equation}
\frac{1}{k} = \frac{2}{\alpha}\frac{v}{A} =
  \frac{2(2-q)(4-q)}{3\alpha} \lambda.
\end{equation}
(The symbol $\lambda$ refers to the mean free path parallel to the
magnetic field, $\lambda_\parallel$, unless otherwise specified.)
For $q=1$ and $u/v=0.02$, deviations from the solution given by the
standard diffusive approximation are expected upstream within 
$1/k=0.13\lambda$ of the shock (corresponding to $\alpha_2$) and downstream 
within $0.14\lambda$ (corresponding to $\alpha_{-1}$); for $q=1.5$
those distance scales become $0.07\lambda$ and $0.08\lambda$, respectively.

Note that this analysis alone cannot give the relative amplitudes of
different solutions; for that, one must consider how particles cross
the shock.  Kirk \& Schneider (1987a) presented eigenvalues and eigenfunctions
of equation (\ref{eq:e}) for the case of $q=1$, and for the case of
ultrarelativistic particle energies ($E\approx pc$)
and a parallel shock, were also able to 
match the upstream and downstream solutions and
analytically evaluate the steady-state distribution function, as well as 
the power-law index of the resulting spectrum.
The analogous calculation for oblique shocks was performed by 
Kirk \& Heavens (1989).  For parallel shocks, 
Kirk \& Schneider (1989) relaxed the assumption of ultrarelativistic
particle velocities in calculating the steady-state particle
distribution as a function of position, pitch angle, and momentum.
In this work, we treat the crossing of moderately relativistic 
particles across an oblique
shock in numerical (finite difference) solutions of the pitch angle transport
equation, and we also compare our steady-state numerical 
results with pitch angle distributions and spatial decay constants,
$k$, that are expected based on the eigenfunctions and eigenvalues of
equation (\ref{eq:e}), respectively.

In summary, the results for steady-state shock acceleration can be
expressed in terms of a superposition of separable solutions of
equation~(3).  A numerical solution of equation~(1), in our case by a
finite difference method, is required in order to determine the relative 
amplitudes of the separable solutions.  We stress that the numerical
method does not explicitly involve the eigenvalues and
eigenfunctions, so we can test the numerical code by verifying that
the results are consistent with a linear combination of separable
solutions.

\subsection{Fluid and Magnetic Field Parameters}

Based on equations (133)-(135) and (124) of de Hoffmann \& Teller (1950),
the jump conditions of a non-relativistic shock in the de Hoffmann-Teller 
frame are
\begin{eqnarray}
\left[u_n + \frac{3}{5}\frac{u_s^2}{u_n}
 + \frac{u_{An}^2t^2}{2u_n}\right] &=& 0 
 \nonumber \\
\left[\left(u_n-\frac{u_{An}^2}{u_n}\right)t\right] &=& 0 
 \nonumber \\
\left[\frac{3}{2}u_s^2 + \frac{1}{2}u_n^2(1+t^2)\right] &=& 0 
 \nonumber \\
\left[\frac{u_{An}}{\sqrt{u_n}}\right] &=& 0,
\label{eq:dHT}
\end{eqnarray}
where a bracketed quantity refers to the difference between that quantity on 
either side of the shock, $u_s=\sqrt{(5/3)(p/\rho)}$ is the speed of sound, 
$u_{An}\equiv B_n/\sqrt{4\pi\rho}$ is the Alfv\'en speed corresponding
to $B_n$ (hence the Alfv\'en speed is $u_A=u_{Az}\sec\theta$), 
and $t=\tan\theta$, where $\theta$ is the
angle between the magnetic field and the shock normal.
In the above, we used the conservation of mass flux and assumed an ideal gas 
with an adiabatic index of 5/3.  Analogous conditions for relativistic 
shocks were derived by Heavens \& Drury (1988).

Equations (\ref{eq:dHT}) relate $u_n$,
$u_s$, $u_A$, and $t$ upstream and downstream of the shock, so given
any four of those variables the remaining four can be determined.
We use the subscript 1 (2) to refer to upstream (downstream) variables.
Solutions were obtained for $u_{s1}=u_{A1}=50$ km s$^{-1}$, typical of 
near-Earth interplanetary space, and selected values of $t_1$ and
$u_{1n}-u_{2n}$.
Note that in the high speed limit $u_{1n}/u_{2n}\rightarrow 4$, i.e.,
$(\gamma+1)/(\gamma-1)$ for an adiabatic index $\gamma=5/3$.  In this limit, 
the magnetic field tension becomes negligible, and the tangential component 
of the velocity, $u_n t$, is nearly conserved, so we have 
$t_2/t_1\rightarrow 4$ as well.

The only fluid parameters that directly affect the particle transport 
are the fluid speed upstream and downstream and the magnetic compression 
ratio, $B_2/B_1$.  Since the normal component of $\vec B$ is conserved,
we have $B_2/B_1=\sec\theta_2/\sec\theta_1$.  Figure \ref{f:bcomp} shows the 
shock compression ratio, $u_{1n}/u_{2n}$, and the magnetic compression ratio, 
$B_2/B_1$, as a function of $\Delta u_n=u_{1n}-u_{2n}$ for selected values of
$t_1=\tan\theta_1$.  Since $t_2/t_1\rightarrow 4$ in the limit of
high Mach numbers, the asymptotic value of the magnetic compression ratio is 
$\sqrt{(1+16t_1^2)/(1+t_1^2)}$, which is less than 4.  
To examine the effect of $B_2/B_1$ on cosmic 
ray transport and acceleration near the shock, we perform simulations
for $u_{1n}-u_{2n}=400$ km s$^{-1}$ and $t_1=0$, 1, and 4,
corresponding to upstream shock angles of 0$^\circ$, 45$^\circ$, and
76$^\circ$ and downstream shock angles of 0$^\circ$, 76$^\circ$, and
86$^\circ$, respectively.  Complete fluid and magnetic field 
parameters for these configurations are given in Table 1.
We do not consider $t>4$ because we expect that our model would be
inaccurate for nearly perpendicular shocks in that it neglects particle 
diffusion perpendicular to the magnetic field, which is important for 
highly oblique shocks (e.g., \cite{jjb98}).

\placefigure{f:bcomp}
\placetable{t:config}

\subsection{Boundary Conditions}

Next we consider changes in the momentum and pitch angle as a charged 
particle encounters the shock, i.e., a sudden change in the magnetic field 
and fluid speed.  Following Decker (1983), we assume that in the shock frame 
both the particle momentum and the first adiabatic invariant, i.e.,
the magnetic moment $p_\perp^2/(2meB)$, are conserved.  The above assumption 
is tantamount to 
considering the adiabatic limit, in which the shock is a region where 
the pitch angle changes gradually.  The opposite limit would consider an 
infinitely thin shock front, and in this case the pitch angle distribution 
can only be determined by computing particle orbits.  For highly oblique 
shocks ($\theta\gtrsim 80^\circ$), there is essentially no difference 
between the resulting pitch angle distributions in the two limits 
(\cite{t79}).  Therefore, the approximation we are using should be reliable 
for highly oblique shocks, and at least represents a well-defined limit for 
less oblique shocks.  

The original simulation code of Ruffolo (1995) is capable of treating 
multiple values of the
momentum in order to examine the particle distribution as a function
of the momentum.  However, for simplicity we have only treated one
value of the momentum in this work, and have assumed that the
dependence of $F(t,\mu,z,p)$ on $p$ is given by a power law,
$F\propto p^{-\gamma}$.  When particles are accelerated during
an encounter with the shock, we consider them to be advected from a
lower value of the momentum to the momentum of interest.  Therefore,
we should evaluate the accuracy of the assumption
that $F$ has the same dependence on $\mu$ and $z$ at
the lower values of $p$ from which particles are advected.  
For non-relativistic particles with $u<v$, 
the largest fractional momentum change that can be achieved (that for
reflection back upstream) is $<2u_1/v$ (see Figure 2).  
In Figure 3 we see that the dependence of the eigenvalues $\alpha$ on 
$u/v$ is linear or weaker for $u/v$ values of interest.
As will be described in \S 3.1, we do in fact see a systematic error in
some results that increases with $u_1/v$.
For ultrarelativistic particle speeds (e.g., \cite{ks87a}) the
assumption of a power-law dependence in the steady state is not a problem
because $v\approx c$ so that $u/v$ does not vary with momentum.

Now we turn to the boundary conditions at the edges of the simulation region.
At a reasonable distance from the shock front, we can employ the 
diffusion approximation, in which equation (\ref{eq:tr}) can be expressed as 
a diffusion-convection equation.  In the manner of Earl (1974), we set 
$F(t,p,\mu,z)\approx F_0(t,p,z) + F_1(p,\mu,z)$, where 
$F_0\equiv\langle F\rangle_\mu$, $F_1$ is an odd function of $\mu$, and
higher-order even functions are neglected.  By separating the transport 
equation into odd and even parts, integrating each over
$\mu$, and neglecting the time dependence of $F_1$ (i.e., assuming 
$|\pd F_1/\pd t| \ll |v\pd F_0/\pd z|$), we obtain
\begin{equation}
\frac{\pd F_0}{\pd t} = -\frac{\pd S_z}{\pd z},
\label{eq:da}
\end{equation}
where 
\begin{equation}
S_z = uF_0 - D\frac{\pd F_0}{\pd z}
\end{equation}
is the $z$-flux
and the spatial diffusion coefficient, $D$, is given by equation (\ref{eq:d}).
Equation (\ref{eq:da}) should be obeyed on either side of the shock, with 
special jump conditions at the shock as described above.  Note that 
either $F$ or the phase space density, $f$, may be used in these
equations and in the following discussion.

We note further that in terms of $F_1(p,\mu,z)$, we have
\begin{equation}
S_z = u\left(1-\frac{1}{3}{v^2}{c^2}\right)F_0
 + \frac{v}{2}\int_{-1}^1\mu F_1 d\mu,
\end{equation}
and if $F_1=\delta\mu$, so that $\delta$ is the dipole anisotropy in the 
mixed frame, we obtain 
\begin{equation}
S_z = uF_0 + \frac{v}{3}\left(\delta-\frac{uv}{c^2}\right)F_0,
\end{equation}
where $\delta-uv/c^2$ represents the dipole anisotropy of $F_w$, the 
distribution function defined entirely in terms of independent variables 
in the solar wind frame.

In the diffusion approximation solution of Krymskii (1977), there are
boundary conditions that $F$ remain finite as $z\rightarrow\pm\infty$.  
In the steady state, 
equation~(\ref{eq:da}) requires that $S_z$ be constant on either side of the 
shock, so we have
\begin{equation}
F_0=A + B\exp\left(\frac{u}{D}z\right)
\end{equation}
with different constants $A$ and $B$ upstream or downstream.  To
avoid divergence far downstream, we
must have $B=0$, so $F_0$ is constant, with $S_z=uF_0$.  
Upstream, we have $S_z=uF_u$, where $F_u$ is $F_0$ far upstream of
the shock, representing the existing particle population that is
accelerated by the shock.

In the numerical solution of equation~(\ref{eq:tr}) in terms of $z$ we can
only treat a finite domain.  However, the diffusion approximation is
valid sufficiently far from the shock (see \S 2.3), with $\pd S_z/\pd z = 0$,
so setting $S_z$ at the boundaries of the simulation region is
approximately equivalent to fixing its value at $\pm\infty$.
At the upstream boundary, we set $S_z$ to a constant (for
a given $p$) that is interpreted as $uF_u$.
At the downstream boundary, $S_z$ is set to $u\langle F\rangle_\mu$ 
at the boundary.  In the diffusion approximation (eq.~[12]), this implies 
that $\partial F_0/\partial z=0$, and also that the anisotropy is given by
$\delta=uv/c^2$; the distribution function in the wind frame is
$F_w=(1-\mu uv/c^2)F$ and thus has a dipole anisotropy of zero.

\subsection{Numerical Method}

The numerical simulations reported here employed the finite
difference method of
Ruffolo (1995) as substantially modified by Nutaro, Riyavong, \& Ruffolo (in
preparation).  The latter report will contain details of the
modifications and testing of the new code.  Here, for completeness,
we briefly outline the changes to the treatment of streaming and
convection, and the treatment of particles crossing the shock.
For a description of other aspects of the numerical method which remain 
unchanged, the reader is referred to Ruffolo (1995).

Changes to the treatment of streaming and convection were motivated
by the work of Hatzky (1996), who used the total variation
diminishing (TVD) technique (\cite{s84}).  While our previous
numerical method (Ruffolo 1991, 1995) eliminated spatial differencing and
hence numerical diffusion in the evaluation of these terms, it required a
small step size, $\Delta z=\Delta\mu v\Delta t$.  On the other hand,
the TVD algorithm permits 
only a very small amount of numerical diffusion for significantly larger
$\Delta z$.  The result is that the new code runs 1 to 2 orders of
magnitude faster.  

Another benefit is that convection is now treated in each step instead of by
occasional jumps as in our previous method, yielding much
smoother profiles without spatial averaging.  The previous method was
adequate for the transport of solar energetic particles, where one
could average over a moderate spatial interval, and the
jumps merely yielded small-amplitude oscillations in the intensity as
a function of time.  However, for the present simulations of 
steady-state shock acceleration, with multiple reflections from the
boundaries and a requirement of fine spatial resolution, the new treatment 
of streaming and convection was essential in obtaining the smooth
profiles shown here.  The results for steady-state particle acceleration
at a parallel shock provide a sensitive test confirming the 
accurate treatment of streaming and convection.

In our implementation, we modified the standard TVD algorithm 
in order to permit a Courant number, $\gamma=v_z\Delta t/\Delta z$,
greater than one.  This permits an arbitrary $\Delta z$, and in
particular, for $\Delta z=\Delta\mu\,v\Delta t$ we were able to
reproduce results for solar energetic particles that used our previous 
code (\cite{rk95}).

The other modification of the code was to treat how particles cross
the shock, which was also part of the streaming/convection step.
For each $\mu$-$z$ grid point, we determine whether particles will
stream/convect as far as the shock.  If so, we perform 
a Lorentz transformation into the shock frame, allow particles to
cross or reflect while conserving the magnetic moment, and perform
another Lorentz transformation back into the local wind frame (see
\S2.1).  The TVD algorithm effectively splits a cell into fractions
of particles destined to move to two different spatial locations; at
the shock we generalize this approach to split cells into fractions
destined to move to different $\mu$-cells as well.  The accuracy of 
this technique is verified by the numerical results for
steady-state shock acceleration at oblique shocks, which are
consistent with a sum of separable solutions of the transport
equation (\S2.3) on either side of the shock.

\section{Results}

\subsection{Steady-State Shock Acceleration}

In a steady state, an equilibrium is reached in the evolution
of the particle distribution function in terms of position, pitch
angle, and momentum.  In the present simulations, we assume that the
momentum dependence is given by $F\propto p^{-\gamma}$, so we find
the value of $\gamma$ that yields a steady solution for $F$ in terms
of $z$ and $\mu$.  Figure 6 indicates the flux balance that
determines $\gamma$.  Well away from the shock, we can use the 
diffusion approximation (see \S 2.5) to say that the net $z$-flux 
far downstream is due to convection: 
$S_z=u_2F_0$, where $F_0=\langle F\rangle_\mu$.
Far upstream, there is a balance between convection toward the shock
and diffusion away, so that $S_z=u_1F_u$, where $F_u$ is the far
upstream flux.  Here we set $F_u=0$, so there is a net outflow of
particles from the shock in the downstream direction.  This is
balanced by the $p$-flux, $S_p$, representing 
acceleration at the shock of particles from lower momenta to the momentum of
interest, as well as from the momentum of interest to higher momenta,
for the appropriate steady-state power-law index, $\gamma$.

\placefigure{f:arrows}

Before considering oblique shocks, we tested our methodology for 
the case of a parallel shock ($\theta_1=\theta_2=0$).  Fluid and
magnetic field parameters were as listed in Table 1.  We considered
protons with speeds $v=0.5$ and $0.1 c$, corresponding to kinetic
energies of 145 and 4.7 MeV and momenta of 541.7 and 94.3 MeV
$c^{-1}$, respectively.  We used grid
spacings of $\Delta z/\lambda = 0.025$ and 0.005, respectively, 
and $\Delta\mu = 2/95$.  For convenience, we set $v\Delta t=\Delta z$.
Outer boundaries were placed at $\pm2.5\lambda$ and $\pm0.5\lambda$
for $v=0.5c$ and $0.1c$, respectively.  The numerical treatment of
the boundary conditions 
assumed $F_u=0$ and was sufficiently accurate that the proper behavior of 
$F$ in the diffusion approximation was maintained out to within a few grid 
points from the boundaries, and the location of the boundaries did not
influence $F$ near the shock.  A full simulation required about 2
hours of CPU time on a Sun Ultra-1 workstation.

Figure 7 (top panel) shows the spatial dependence of the pitch angle 
averaged phase space distribution, $\langle f\rangle_\mu$, near a
parallel shock for $v=0.5c$ and $q=1$ ($z>0$ is the upstream region).  
As explained in \S 2.2, our simulations solve for 
$F=d^3N/(dzd\mu dp)$, the density of particles in a flux tube, which
is related to $f$ by $F=2\pi ap^2f$, where $a$ is the cross-sectional
area of a flux tube; therefore, $f\propto BF$.  For a parallel shock,
$B$ is equal on the upstream and downstream side, so $f\propto F$.
Throughout this section, $f$ is normalized to 1 far downstream.
{}From Figure 7 we see that for a parallel shock, the spatial
dependence of $\langle f\rangle_\mu$ is simply that expected in the
diffusion approximation (\S 2.5), i.e., constant downstream and 
exponentially decaying to zero upstream with $k=u/D$.  
The pitch angle dependence was also the same as in the diffusion
approximation.
The same results were obtained for $q=1.5$ and for $v=0.1c$ with $q=1$ and 
1.5.

\placefigure{f:ssfz}

For a coarser $\mu$-grid spacing, a spurious peak was
obtained in $\langle f\rangle_\mu$, 
which tends toward zero as $\Delta\mu\rightarrow 0$ (a
remnant of this is barely visible in the top panel of Figure 7).  
Even for a fine grid spacing, one anomaly is that a value of
$\gamma=2.020$ was needed for a steady state (i.e., to conserve the flux of
particles in the simulation region) for $v/c=0.5$ or 0.1 and $q=1$ or
1.5, whereas acceleration theory for
parallel shocks (e.g., \cite{k77}) yields $\gamma=3u_2/(u_1-u_2)+1$
or 2.034 for these fluid parameters.  The assumptions of that theory
should apply given that $f(\mu,z)$ is that expected from the
diffusion approximation.  We believe that the systematic error in
$\gamma$ obtained from the simulations arises from the assumption of
a power-law dependence of $F$ on $p$, which neglects the momentum
dependence of $F(\mu,z)$ (see \S 2.5).  The key
problem is that the upstream anisotropy of $F$ depends on $u_1/v$.
As a test, the code was modified to artificially add an ``extra''
anisotropy for lower momenta, i.e., to multiply $F$ upon
acceleration by a $\mu$-dependent factor to account for the
higher anisotropy of $F$ at the lower momentum from which a particle
was accelerated, yielding a similar $F(\mu,z)$ and 
$\gamma=2.041$.  We conclude that this explanation 
can in fact account for a systematic error in $\gamma$ of the observed
magnitude, and that $\gamma$ is more sensitive to the assumption of a
power-law dependence than is the distribution of particles in space or
pitch angle.

Turning to oblique shocks, Figure 7 shows the spatial dependence of
$\langle f\rangle_\mu$ for $\tan\theta_1=1$ and 4 and for $q=1$ and
1.5.  In all cases, the distribution function farther from the shock is
consistent with the diffusion approximation, with $\langle
F\rangle_\mu$ constant downstream and exponentially decaying upstream
with $k=u/D$.  A conspicuous feature of Figure 7 is the jump
(discontinuity) in 
$\langle f\rangle_\mu$ at an oblique shock (the finite slope is due
to the finite grid spacing in $z$).  This feature was also found
in simulations by Ostrowski (1991), Gieseler et al.\ (1998),
and T.\ Naito (private communication, 1998).
Gieseler et al.\ (1998) present a detailed theoretical
and computational analysis of this feature, as well as possible
observational signatures.
We find that the jump is stronger for more oblique shocks, 
and weaker for $q=1.5$
than for $q=1$.  The amplitude of the jump is on the
order of a few percent for such fast particles ($v=0.5c$), and
our simulations indicate that the jump is stronger for 
slower particles ($v=0.1c$), i.e., a higher $u/v$.

Another difference from the case of a parallel shock is
that for oblique shocks,
additional eigenfunctions are excited in $f(\mu,z)$ near the shock.
(If one is not sufficiently careful in treating the boundary
conditions, as I was not 
during the initial stages of this work, additional eigenfunctions are also 
excited near the boundaries; discretization errors also yield
spurious eigenfunctions near the shock, which become negligible for 
95 $\mu$-grid points as used here.)  
For all steady-state simulations, $f(\mu,z)$ was consistent with a 
sum of separable solutions of equation (3).
For $\tan\theta_1=4$ ($\theta_1=75^\circ$), Figure 8 shows the
dependence of $f$ on $\mu$ and $z$ within $\pm0.8\lambda$ of the shock, and
Figure 9 shows $f$ as a function of $\mu$ for $z=\pm0.05\lambda$.
(Recall that we use $\mu$ and $p$ to refer to quantities in the local
fluid frame; thus these plots are for a constant value of the local $p$.  
A Compton-Getting transformation to the shock frame would 
have no noticeable effect on our distribution plots.)
For $\tan\theta_1=1$, the results were qualitatively similar but with
weaker anisotropies.

\placefigure{f:ssfmuz}
\placefigure{f:ssfmu}

In Figure 9, we see that 
upstream distributions (thick lines) increase with $\mu$ up to 
$\mu\approx0.7$ (with a slightly stronger anisotropy than in the far 
upstream region), and for greater $\mu$ values, $f$ drops sharply.
The reason for the sharp drop is that given our assumption of
conservation of the magnetic moment, particles with
$\mu>\sqrt{1-B_1/B_2}$, or 0.85 in this case, have come from
downstream.  A similar drop in $f$ has been called a ``deficit cone''
(\cite{nea92}) or ``loss cone'' effect (\cite{be97}) for the case of 
galactic cosmic ray (GCR) depletion at high $\mu$ upstream of an 
interplanetary shock, which is due to the paucity of GCR coming from 
downstream (see \S 3.2).  
The same effect occurs here because the acceleration of particles
coming from downstream is weaker than for particles reflected from
upstream.  The greatest acceleration occurs for particles
reflected with the greatest change in pitch angle (see Figure
2), i.e., for $\mu$ slightly below 0.85.  Since stronger acceleration
implies that $f$ is advected from lower momenta, and the particle
spectrum increases with decreasing momentum in this case, the
strongest acceleration corresponds to the greatest increase in $f$.

In the downstream region, particles are redistributed in pitch angle 
because of changes in pitch angle as particles cross the shock; the
average flux also increases slightly due to acceleration.  It is
worth noting that for a highly oblique shock, most particles coming
from upstream are in fact reflected, i.e., when
$|\mu|<\sqrt{1-B_1/B_2}$, or in the case of a strong, highly oblique
shock, for pitch angles more than 30$^\circ$ from the magnetic field
direction.  Another feature of Figures 8 and 9 is the sharp gradient
in $f$ at $\mu=0$ for the case of $q=1.5$.  For this form of the
pitch angle diffusion coefficient, $\varphi(\mu)=A|\mu|^{0.5}(1-\mu^2)$ 
tends to zero as $\mu\rightarrow 0$.  Since the $\mu$-flux, 
$S_\mu=-(\varphi/2)(\pd F/\pd\mu)$, is slowly varying in a
near-equilibrium situation, the vanishing diffusion coefficient at
$\mu=0$ is able to sustain an infinite gradient in $F$ at that value.

We believe that this behavior of $f$ as a function of $z$ and $\mu$
is not an artifact of the assumption of a power-law momentum
dependence because when an extra anisotropy was artificially added,
the $\mu$ and $z$ dependence (including the jump at $z=0$) was not
significantly affected; this was also the case for parallel shocks.  
On the other hand, computed values of $\gamma$ are 
strongly affected by the power-law assumption, so that this code
in its present form is essentially unable to determine $\gamma$.
The error in $\gamma$ was much weaker for $v/c=0.5$ than for
$v/c=0.1$ (because of the lower $u_1/v$ ratio).
As an example, for $q=1$ the $\gamma$ values required for a steady state 
with $v=0.5c$ were 1.965 and 1.952 for $\tan\theta_1=1$ and 4,
respectively, while with $v=0.1c$ they were 1.985 and 1.787, respectively.
Otherwise, the results regarding $f(\mu,z)$ 
for $v/c=0.1$ were qualitatively similar to those shown in Figures
7 to 9 for $v/c=0.5$, with much stronger anisotropies and jumps in
$\langle f\rangle_\mu$ at the shock.

\subsection{Precursors of Forbush Decreases}

To demonstrate the versatility of this method, we apply it to
model Forbush decreases of galactic cosmic rays (GCR)
as an interplanetary shock passes the Earth (\cite{f38,bh42,fl42}), which 
represent a transient phenomenon instead of a steady state.
Ground-based neutron monitors measure secondary neutrons 
from the impact of relativistic, primary charged particles, 
mainly protons, on the upper atmosphere.  Due to selective deflection by
the Earth's magnetic field, neutron monitor observations
are sensitive to primary cosmic rays from specific directions in 
space, and the worldwide network of neutron monitors provides
detailed information on their pitch angle distribution, sensitive to
variations on the order of 0.1\%.  Precursors to Forbush decreases are of 
practical interest as possible predictors of space weather effects on the 
Earth, such as satellite failures, radio fade-outs, power outages, etc., 
several hours or even days before the passage of a major interplanetary
shock.  Several analyses of neutron monitor observations have indicated two 
types of precursors to Forbush decreases: 1) an enhanced diurnal 
anisotropy of GCR, with an excess of particles traveling toward 
the Sun along the interplanetary magnetic field, and 2) a deficit of GCR in
a ``loss cone," i.e., along a narrow range of pitch angles
directed nearly along the interplanetary magnetic field away from the
Sun (\cite{nea92,nfm94,sea95,bea95,crv96,be97}).  

We model the Forbush decrease in a rather idealized
manner, assuming the configuration of Figure 1 and 
neglecting particle drifts (\cite{n83,kn86}), spatial
dependence of the scattering mean free path, shock curvature,
the finite spatial extent of the interplanetary shock, adiabatic
focusing, and adiabatic deceleration.  Nevertheless, we can
explain the basic features of the observed precursors, verifying
their interpretation in terms of particle transport in the vicinity of 
an oblique shock.

The simulation conditions were inspired by the dramatic CME event of 1997 
April 7, which arrived near Earth on April 10-11, and for which a
possible loss cone feature is identified by Bieber \& Evenson (1997).  
In this case, the travel time of 3 days indicates a shock speed of 
$\approx 600$ km s$^{-1}$ or only 200 km s$^{-1}$ faster than the (typical)
solar wind speed.  Assuming that the shock normal is radial,
we take the upstream shock-field angle to be the typical ``garden-hose''
angle of 45$^\circ$ ($\tan\theta_1=1$), and as before we assume
$u_{s1}=u_{A1}=50$ km s$^{-1}$.  Thus we find $\Delta u_n=133$ km s$^{-1}$,
$\tan\theta_2=3.20$, and $B_2/B_1=2.37$, which in turn
implies that particles crossing the shock from downstream
have pitch angles aligned with the magnetic field to within
40$^\circ$ ($\mu>0.76$).  We used $q=1.5$, which adequately
describes interplanetary scattering (Bieber et al.\ 1986). 
For the upstream boundary condition, we specify a constant $F_u$ 
(see \S 2.5), and the initial condition sets $F$ to that constant in the 
upstream region and to zero in the downstream region.  We used 
$\Delta\mu=2/45$ (45 $\mu$-grid points) and $\Delta z/\lambda=0.05$, where 
$\lambda=0.3$ AU.  We considered $v=0.75c$, corresponding
to a kinetic energy of 480 MeV.  For the momentum spectrum, we
assumed $F\propto p^{-1}$, according to the model proton spectrum of
Reinecke, Moraal, \& McDonald (1996) for the similar polarity solar cycle 
of 1977.  However, the simulation results were very insensitive to the GCR 
spectral index.

Figure 10 shows the omnidirectional GCR intensity as a function of
position (normalized to 1 just upstream of the shock).  As the shock moves 
past a fixed observer, one sees a
gradual precursor decline and a slight recovery as the shock
approaches.  Such gradual declines in some observations were noted by
Cane et al.\ (1996).  In their one figure for such a Forbush
decrease, that of 1972 Oct 31 (Figure 5 of that paper), it is seen that 2
out of 3 neutron monitor stations observed a relative peak near the
time of shock onset.  It would be interesting if the omnidirectional
flux could be estimated from the worldwide neutron monitor network
for such events for direct comparison with the results of numerical 
simulations.

\placefigure{f:gcrfz}

At the shock itself, we see a jump reminiscent of that found in \S
3.1 for shock acceleration with $F_u=0$ (Figure 7; see also
\cite{o91,gea98}).  Thus this model predicts a discontinuous drop at
the shock followed by a more gradual decline which we identify as the
declining slope of a Forbush decrease.  We note, however, that this
model cannot hope to accurately reproduce the detailed features of
the Forbush decrease itself, given its simplistic assumptions (see
also \S 4).

The phase space distribution, $f$, as a function of $\mu$ and $z$ is
shown in Figure 11, and pitch angle distributions in the near
upstream ($z=0.05\lambda$) and far upstream ($z=\lambda$) regions are
shown in Figure 12.  We see that there is an overall enhanced sunward
anisotropy both downstream and upstream, where it persists over
several mean free paths from the shock.  Close to the shock, we also
find the second type of observed precursor feature with a sharp
decline in the loss cone region.

\placefigure{f:gcrfmuz}
\placefigure{f:gcrfmu}

Our numerical simulations
indicate that $f(\mu,z)$ upstream is approximately given
by a sum of steady-state separable solutions.  If a steady
state has not been achieved, separable solutions of the full transport 
equation involve an eigenvalue equation similar to equation (5) with
$u/v\rightarrow u/v-1/(kv\tau)$, where $\tau$ is the decay time. 
Given the weak sensitivity of eigenvalues to $u/v$ (except
$\alpha_1$), the longest-lived solutions should be similar to the
steady-state solutions of equation (5) with some modification to $k$.  
The time scale of evolution of a Forbush decrease, $\sim$1 day, is much 
longer than the travel time of particles across the distance scales of
interest, $\sim$10 minutes, so it is reasonable that transient solutions
have disappeared leaving only nearly stationary solutions of the 
transport equation.

Both types of observed precursors of Forbush decreases
are readily understood in terms of 
the upstream simulation results, as expressed as a superposition of
separable solutions of equation (3).  The constant solution in the far
upstream region corresponds to the zero eigenvalue.  The 
contribution of the first eigenfunction is opposite in sign to
that seen in \S 3.1, corresponding to an anisotropy directed toward the
Sun over a long spatial scale.  Turning to the next eigenfunction, 
corresponding to $\alpha_2$, the spatial decay scale of the
steady-state eigenfunction is 0.074$\lambda_\parallel$, and in the 
time-dependent simulations this feature has a spatial scale length of
$\approx0.08\lambda_\parallel$ along the magnetic field, or
$\approx0.11\lambda_r$ in the radial direction.
This eigenfunction appears with the same sign as in Figure 4 for
shock acceleration (Figure 9) and represents a deficit of particles
in the loss cone ($\mu\gtrsim 0.76$) that came from the downstream region 
which is depleted in GCR, as well as an enhancement of particles with $\mu$
just below 0.76 which were accelerated during reflection from the shock.
Several examples of loss-cone deficits were given by Nagashima et al.\ 
(1994) and Sakakibara et al.\ (1995), and an enhancement for certain
pitch angles corresponding to reflection and acceleration at the shock
was reported by Belov et al.\ (1995).  The predicted spatial decay length
for these features depends only weakly on the shock speed or obliquity
(i.e., the eigenvalues depend only weakly on $u_1/v$, especially for 
relativistic velocities), but there is a 
significant dependence on $q$ (see eq.\ [9]).

The results shown here, with precursor features roughly of the
magnitude reported by Cane et al.\ (1996), were for a simulated
duration of 7.4 hours, which is substantially shorter than the actual
time it takes a shock to propagate from the Sun to the Earth
(typically 2 to 4 days).  This is probably due to our neglect of
enhanced spatial diffusion near the shock, which must be present in order to
account for the sharp GCR gradient at the onset of a Forbush
decrease.  Stronger scattering would stem the tide of equilibration
and lead to a longer duration of these features.

\section{Discussion}

The numerical simulation procedure developed in this work is the first to 
solve a pitch angle transport equation on both sides of an oblique,
non-relativistic shock without assuming an
ultrarelativistic particle velocity.  Here we have shown applications of the 
technique to study oblique shock acceleration in the steady state,
and to consider the time-dependent problem of Forbush decreases and
in particular their precursors.  To explore the capabilities and
limitations of this type of solution, we have started with the
simplest case of a plane-parallel, oblique shock with straight
magnetic field lines on either side (Figure 1) and a spatially
uniform scattering mean free path.  Clearly this simple model is
neglecting a variety of important processes, which will be discussed
shortly.

One limitation in this work was our assumption of a power-law
dependence of the distribution function on the particle momentum.  In
further work this assumption should be relaxed, i.e., the simulations
should treat different values of the momentum, as has been done
by Kirk \& Schneider (1989).  This assumption has strongly
affected calculated values of the particle spectrum, but based on test
runs that artificially compensate for the error, it seems not to
have significantly affected the spatial and pitch angle
distributions.  This should be checked in future work.  Monte Carlo
(MC) techniques are undoubtedly more popular than the finite
difference method used here, and MC simulations are very well-suited
for exploring certain phenomena.  Here we use a finite difference
method for consistency with an existing code that includes a variety of 
effects known to be important for interplanetary transport (Ruffolo
1995) so that the treatment of the shock can be readily incorporated
into such simulations.  For transport simulations in general, finite 
difference or eigenfunction expansion techniques have some advantages, 
such as often requiring less computing time, the absence of statistical 
error, straightforward extrapolation to reduce discretization error, 
easier treatment of a high dynamic range in particle density, and a
straightforward and continuous implementation of analytic expressions for
scattering and other processes
(for a comparison of different techniques, see \cite{eea95}).
The last of these is also a notable limitation in situations where
analytic expressions are not available, e.g., situations requiring
the tracing of particle orbits.  If the tracing of particle
orbits is only necessary near the shock, that could be implemented in 
a finite difference code by a ``transfer matrix'' based on a one-time
evaluation of the probability of various outcomes at the shock
for each $\mu$-grid point.

To the author's knowledge, calculated pitch angle distributions for 
steady-state acceleration of energetic particles
at a non-relativistic oblique shock have not reported previously.  
Such pitch angle distributions should be directly comparable with recent and
upcoming {\it in situ} observations near interplanetary shocks.  As observed
distributions are reported in greater detail, they can be expected to
challenge the theoretical models and help indicate which additional
physical processes have important effects on particle
transport and acceleration at interplanetary shocks.

The pitch angle distributions for steady-state shock
acceleration represent the superposition of
pitch angle eigenfunctions with different amplitudes, as first
described by Kirk \& Schneider (1987a).  The amplitudes of
different eigenfunctions can only be determined by treating how
particles cross the shock.  The upstream pitch angle distributions obtained
here for oblique, non-relativistic shocks and mildly relativisitic particles
(Figure 9) are qualitatively very different from those found for 
parallel, relativistic shocks and ultrarelativistic particles 
(Kirk \& Schneider 1987a,b, \cite{hd88}).  In contrast to the
non-relativistic case, those authors found that even for a parallel
shock, eigenfunctions beyond those present in the diffusion
approximation are excited.  The upstream pitch angle
distribution is highly collimated in the direction away from the
shock.  The downstream distribution for the parallel, mildly
relativistic shock ($u_1=0.3c$) is qualitatively similar to 
what we find for oblique, non-relativistic shocks.
Our results can be more directly compared with the pitch angle
scattering results of Naito \& Takahara (1995), who treated oblique, mildly 
relativistic shocks with $u_1=0.1c\cdot\sec\theta_1$.  We find qualitatively 
similar upstream pitch angle distributions but very different downstream
distributions; for high obliquity (and high $u_1$) those of Naito \& 
Takahara (1995) are strongly collimated with particles moving away from the 
shock.

The second application considered here, to precursors of Forbush 
decreases of galactic cosmic rays (GCR), 
demonstrates the ability of this method to simulate
time-dependent phenomena.  The eigenfunction analysis alone is
sufficient to specify that in a steady state, one can have 
certain types of features in the
pitch angle distribution over certain distance scales.  For example,
the observed increase in diurnal anisotropy well before the onset of
some interplanetary shocks (e.g., Cane et al.\ 1996, \cite{be97}) is 
identified with the eigenvalue $\alpha_1$ and a long distance scale,
and the superimposed excess for $\mu$ up to $\approx0.5$ and strong deficit
thereafter (Nagashima et al.\ 1992, 1994, \cite{sea95}) is identified with 
the eigenvalue $\alpha_2$ and a distance scale of $\sim0.1\lambda$.  The
time-dependent simulations are necessary to
verify that these modes are in fact
excited with the appropriate sign and a reasonable amplitude, and
that the time-dependent, upstream 
distribution is approximately represented by
a superposition of steady-state eigenfunctions (which is not the case 
downstream).  

We note that simulations of a Forbush decrease
with a parallel shock did not exhibit features corresponding to
$\alpha_2$, which is an example of how not all eigenfunctions are
excited in every situation.  Since a localized deficit in a narrow loss
cone is in fact not expected for a parallel shock, this supports
the physical explanation that the upstream deficit over a narrow
range of pitch angles corresponds to particles crossing
from the downstream region which is depleted in GCR
(\cite{nea92}).

As stressed earlier, this model of a Forbush decrease is idealized in
many ways, though it seems to be adequate for describing the key
features of upstream precursors, which are of practical interest for
short-term space weather forecasting (\cite{be97}) by warning of the
impending impact of a major interplanetary shock.  We are much more
hesitant to apply this idealized model to the Forbush decrease
itself.  However, according to Wibberenz, Cane, and Richardson (1997), 
the key features of a Forbush decrease can be captured by assuming
enhanced scattering (a lower mean free path) in the region just
downstream of the shock.  This could easily be included in a
simulation technique such as ours and a comparison with particle
distributions observed during the course of Forbush decreases could
help identify what physical processes are crucial to the Forbush
decrease phenomenon.

A key motivation for this work is the potential to include more
physical effects in the future.  For a more realistic magnetic
field configuration, transport effects such as adiabatic
focusing and deceleration can readily be included, as they have
already been included in numerical simulations of interplanetary
transport.  Enhanced scattering near a shock almost certainly affects
acceleration and transport in that region, yet the magnitude and
extent of such scattering for high-energy particles is not well
understood.  Concrete models of the spatial dependence of the mean free 
path near a shock should be developed and tested for their success in
explaining {\it in situ} observations.

\acknowledgments

The author is grateful to Paul Evenson, Wolfgang Dr\"oge, and Marty Lee for
useful discussions and suggestions.
In the formative stages of this work, I greatly benefited from lectures
at the JSPS-ICRR International Spring School in Tokyo during
February, 1995, and especially from those given by Toshio Terasawa.
I particularly want to thank the Thailand Research Fund for their
support of this research.  I am also grateful for the kind hospitality of 
the Bartol Research Institute of the University of Delaware, where part
of this research was performed under the support of NASA grant NAG
5-2606.

\newpage

\figcaption[kfig1.eps]{Model configuration in the de Hoffmann-Teller frame, 
in which the shock is stationary, the fluid flow $\vec u$ is 
parallel to the magnetic field $\vec B$, and the electric field is zero.
\label{f:config}
}

\figcaption{
Illustration of changes in pitch angle when a particle crosses an
oblique shock, a process which contributes to particle acceleration.
Here $v_\parallel$ and $v_\perp$ are components of the particle
velocity in the shock frame.  See text for details.
\label{f:circle}
}

\figcaption{Eigenvalues closest to zero for equation (5)
as a function of $u/v$ for $q=1$ (solid lines) and $q=1.5$ (dashed lines).
\label{f:eigen}
}

\figcaption{Eigenfunctions of equation (5) for eigenvalues
shown in Figure 3 for $u/v=-0.075$ and $q=1$.  For steady-state shock
acceleration, the pitch angle distribution is a linear combination of
such functions.  Particle motion is directed away
from the shock for $\mu<0$ downstream and $\mu>0$ upstream.
\label{f:eigenfunc}
}

\figcaption{Fluid compression ratio, $u_{1n}/u_{2n}$, and magnetic 
compression ratio, $B_2/B_1$, as a function of the normal velocity
jump in the shock frame, $\Delta u_n=u_{1n}-u_{2n}$, for 
$t_1=\tan\theta_1=0$, 0.5, 1, 2, and 4.
\label{f:bcomp}
}

\figcaption{Schematic illustration of the balance of spatial and
momentum fluxes in diffusive shock acceleration.
\label{f:arrows}
}

\figcaption{Spatial dependence of the pitch-angle averaged
phase space distribution function,
$\langle f\rangle_\mu$, for steady-state particle acceleration near a
shock (at $z=0$) for various values of $\tan\theta_1$, the tangent of
the angle between the magnetic field and the shock normal, for $q=1$
(solid lines) and 1.5 (dotted lines), and for $v=0.5c$.  
Farther from the shock,
$\langle f\rangle_\mu$ is constant downstream and exponentially
decays toward zero upstream.
The ordinate is normalized to the value far downstream.
\label{f:ssfz}
}

\figcaption[kfig8.eps]{Phase space distribution of particles
as a function of $\mu$ and $z$
(in units of $\lambda$) near an oblique shock with $\tan\theta_1=4$
for a) $q=1$ and b) $q=1.5$.  Note the changes in the pitch angle
distribution near the shock (at $z=0$).
\label{f:ssfmuz}
}

\figcaption{Phase space distribution of particles
as a function of $\mu$ near an oblique shock with $\tan\theta_1=4$
for $q=1$ (solid lines) and $q=1.5$ (dashed lines) at
$z=0.05\lambda$ (upstream; thick lines) and $z=-0.05\lambda$
(downstream; thin lines).
\label{f:ssfmu}
}

\figcaption{Spatial dependence of the pitch-angle averaged phase
space distribution function, $\langle f\rangle_\mu$, of galactic cosmic
rays near an oblique interplanetary shock ($\tan\theta_1=1$).  As the
shock moves (to the right) past a fixed observer, one sees a gradual
precursor decline and a slight recovery, followed by a discontinuous
decrease at the onset of a Forbush decrease.
\label{f:gcrfz}
}

\figcaption[kfig11.eps]{Phase space distribution of particles
as a function of $\mu$ and $z$ (in units of $\lambda$) near an
oblique interplanetary shock ($\tan\theta_1=1$).  The sharp drop
downstream ($z<0$) represents the Forbush decrease; upstream ($z>0$)
features represent precursors that may be useful for space weather
prediction.  Superimposed on the
enhanced sunward anisotropy, note the unusual pitch angle
distribution just upstream of the shock.
\label{f:gcrfmuz}
}

\figcaption{Phase space distribution of galactic cosmic rays
as a function of $\mu$ near an oblique interplanetary shock
($\tan\theta_1=1$) at $z=0.05\lambda$ (solid line) and $z=\lambda$
(dashed line), representing near upstream and far upstream precursors
of a Forbush decrease.
\label{f:gcrfmu}
}

\begin{deluxetable}{crccc}
\tablecolumns{5}
\tablewidth{0pc}
\tablecaption{
   Selected Solutions of Shock Jump Conditions\tablenotemark{a}
\label{t:config}
}
\tablehead{
\colhead{$\tan\theta_1$} &
\colhead{$\tan\theta_2$} &
\colhead{$u_{1n}$} &
\colhead{$u_{2n}$} &
\colhead{$B_2/B_1$} }
\startdata
 0 & 0\phantom{.00} & 538.0 & 138.0 & 1.00 \\
 1 &  3.88 & 541.5 & 141.5 & 2.83 \\
 4 & 15.11 & 544.3 & 144.3 & 3.67 
\enddata
\tablenotetext{a}{For $u_{A1}=u_{s1}=50$ km s$^{-1}$.
All velocities are in units of km s$^{-1}$.} 
\end{deluxetable}

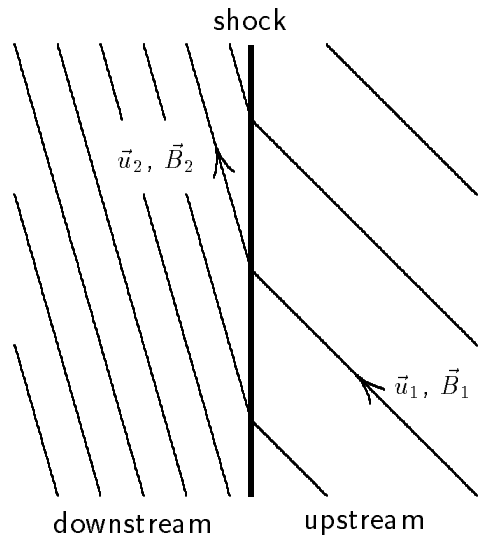
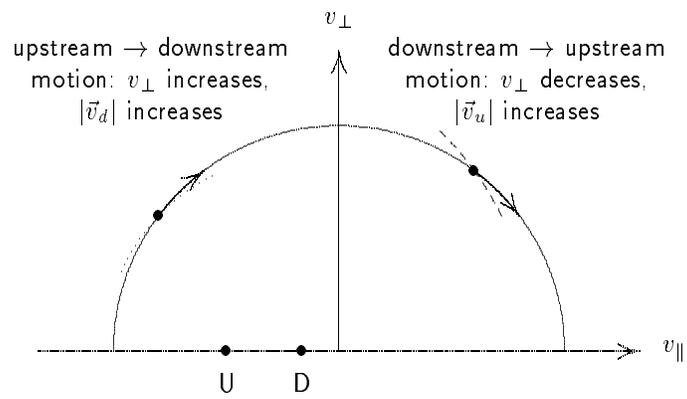
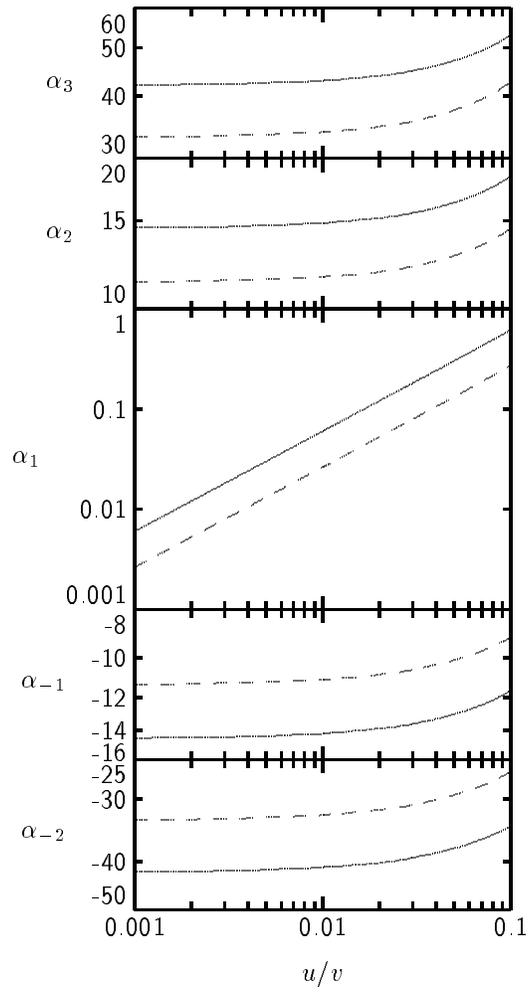
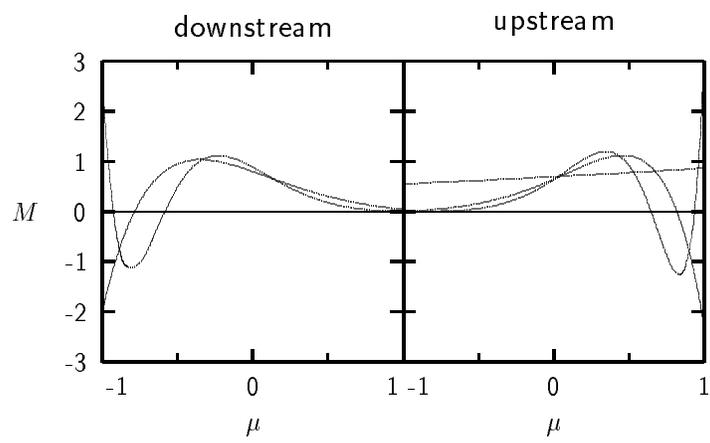
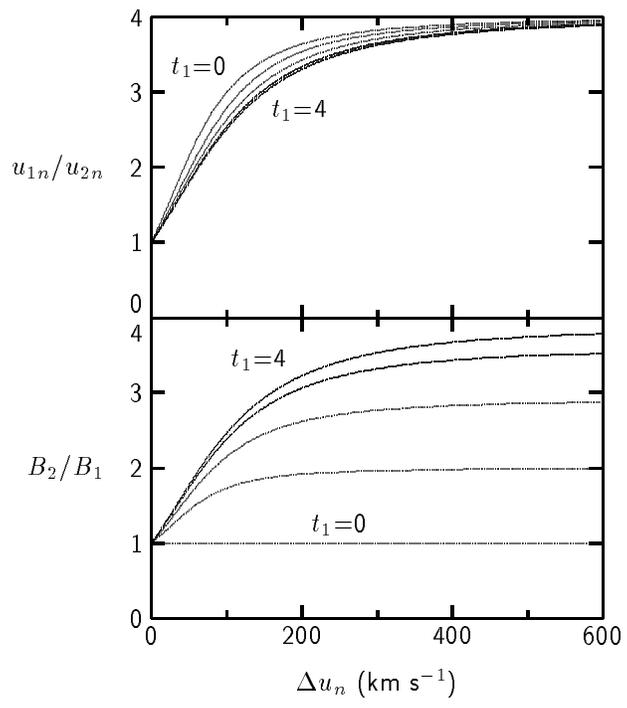
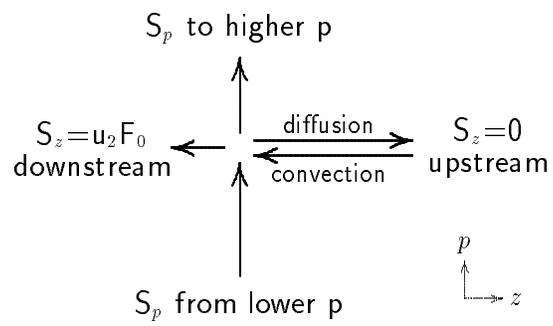
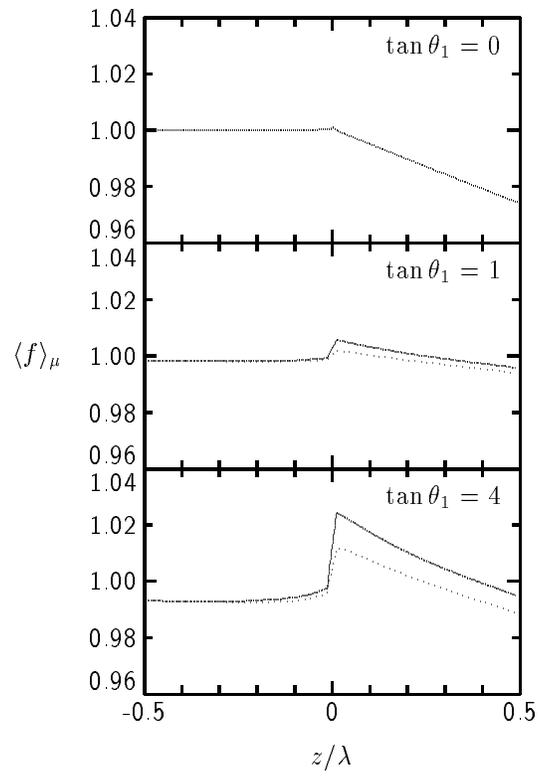
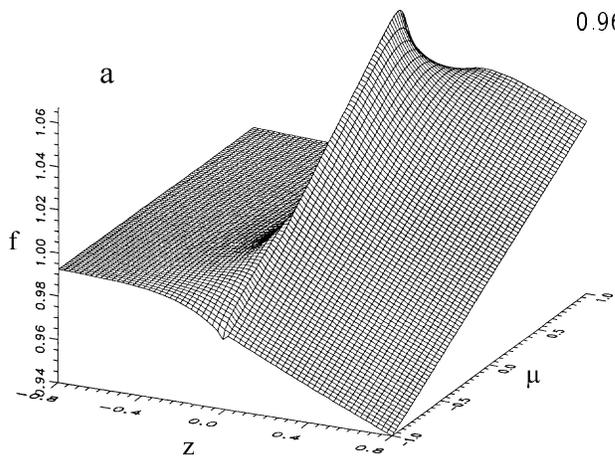
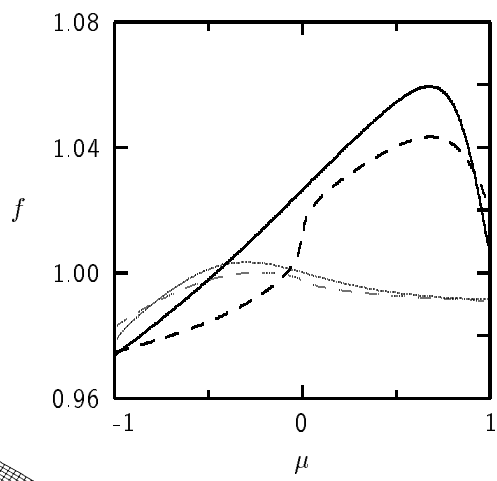
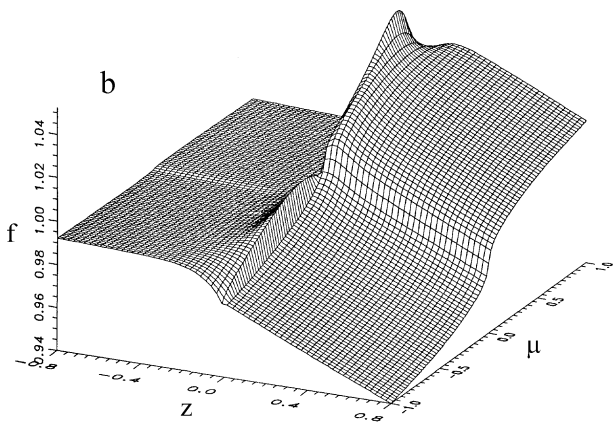
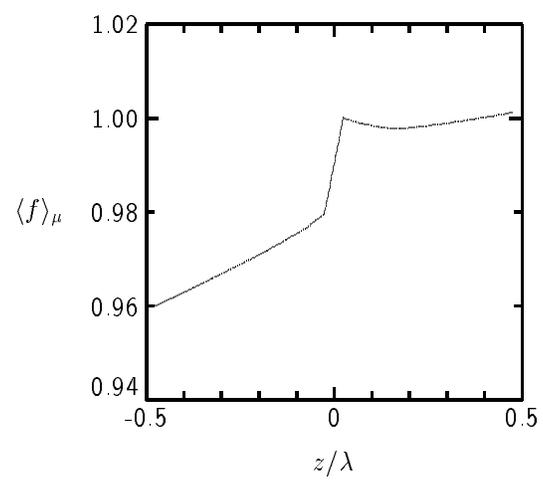
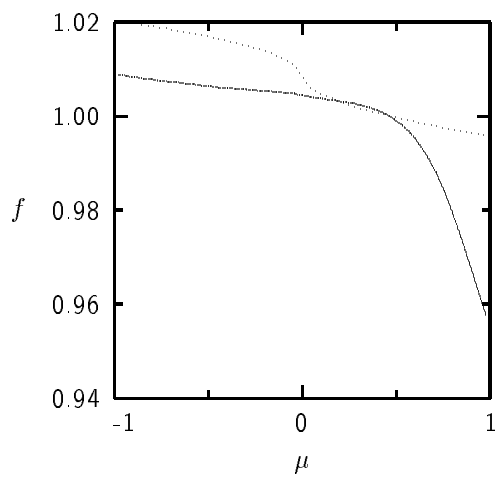
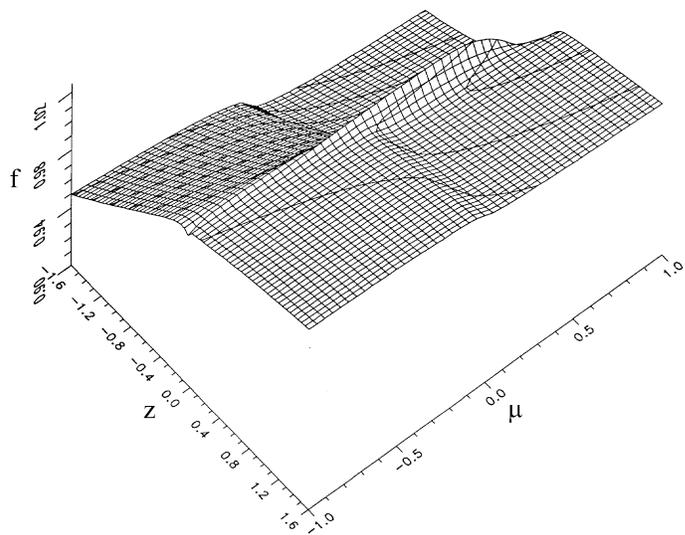
\end{document}